\magnification=\magstep1
\baselineskip=14 truept
\tolerance=10000
\hsize=16truecm
\vsize=24truecm
\centerline {\bf Common Origin of Quantum Regression }
\centerline {\bf and Quantum Fluctuation Dissipation Theorems}
\vskip 0.5truecm\noindent
\centerline{\it P. Shiktorov, E. Starikov, V. Gru\v zinskis}
\vskip 0.2truecm\noindent
\centerline{ Semiconductor Physics Institute,
A. Go\v stauto 11, 2600 Vilnius, Lithuania}
\vskip 0.5truecm\noindent
\centerline{\it L. Reggiani}
\vskip 0.2truecm\noindent
\centerline{ Dipartimento di Ingegneria dell' Innovazione,}
\centerline{  Istituto Nazionale di Fisica della Materia,}
\centerline{ Universit\`a di Lecce, Via Arnesano s/n, 73100 Lecce, Italy}
\vskip 0.5truecm\noindent
\centerline{\it L. Varani and J.C. Vaissi\`ere}
\vskip 0.2truecm\noindent
\centerline{ Centre d'Electronique et de Micro-opto\'electronique 
de Montpellier}
\centerline{ (CNRS UMR 5507), Universit\'e Montpellier II,
34095 Montpellier Cedex 5, France}
\vskip 1.5truecm\noindent

\centerline{\bf Abstract.}
\vskip 0.2truecm\noindent
It is shown that 
the quantum fluctuation dissipation theorem 
can be considered as a 
mathematical formulation in the 
spectral representation of 
Onsager hypothesis on the regression
of fluctuations in physical systems.
It is shown that the quantum fluctuation dissipation theorem 
can be generalized to an arbitrary stationary state.
\vfill\break\noindent
{\it Introduction.}
Under thermal equilibrium conditions 
the  behavior of fluctuations of macroscopic 
observables of a physical system is governed by relationships which  
are formulated usually in terms 
of the regression theorem (the so called Onsager hypothesis [1]) 
and the  fluctuation-dissipation theorem [2-4].
The former pertains to the time domain and
states that the relaxation of a correlation of fluctuations
is described by the same law governing the  irreversible processes of
the observable quantity itself.
The latter pertains to the frequency domain and
interrelates in some universal way
the spectral characteristics of fluctuations and linear response 
(i.e. dissipation) of an observable of the physical system.
In the framework of a classical approach,
the two theorems complement each other providing
a closed description of fluctuations and linear response 
in thermal equilibrium.
By contrast, within the more general quantum approach 
there appears a conflict between 
these two theorems.
This conflict is usually interpreted as a violation
of the quantum regression theorem (QRT) (see, for example, Refs. [5-7]).
In the most evident form such a violation is demonstrated in Ref. [7],
where the conclusion "there is no quantum regression theorem" is announced.
The proof of the general character of such a statement
is based on the fact that the violation of  QRT
follows from the quantum fluctuation dissipation theorem (QFDT).
A proof that QRT is valid independently of  QFDT
was given by Lax [8] on the  basis of
the  general principles of quantum statistics (see also Refs. [9-11]).
Since from Ref. [7] it is claimed that QFDT
contradicts the validity of  QRT, 
we argue  that the origin of such a conflict
is related with QFDT and its interpretation.
\par
The aim of this letter is 
to establish the origin of such a conflict and to provide arguments 
to reconcile the two theorems.
For this sake, the conditions necessary for QRT to be fullfilled
are firstly considered in the framework of the general formalism
of linear response functions.
Then, the equivalence between QFDT and QRT 
is proven by revisiting the derivation of QFDT.
Finally, an extension of QFDT to an arbitrary stationary state
which can be far from that of  thermal equilibrium is presented. 
\par
{\it The Onsager regression theorem.}
Let us define formally a law which will describe the time evolution of 
perturbations of some observable physical quantity $x(t)$.
It is assumed that the response of the observable,
induced by some external perturbation described by a force $f(t)$,
is linear and can be represented as a convolution integral:
$$
x(t)=\int_0^{\infty}\alpha_x(\tau)f(t-\tau)d\tau
\eqno(1)
$$
Here $\alpha_x(\tau)$ is a real function of time determined by the state
and the properties of the physical system under test.
The response described by Eq. (1) satisfies the causality principle,
i.e. $\alpha_x(\tau)=0$ at $\tau < 0$.
Therefore, one always can introduce a linear integro-differential
operator $A_t\{...\}$, such that $x(t)$ given by Eq. (1) is 
a solution of the equation:
$$
A_t\{x(t)\}=f(t)
\eqno(2)
$$
Due to the linearity of  $A_t\{...\}$, 
its eigenfunctions are the harmonic functions $exp(-i\omega t)$, and
the corresponding eigenvalues $A(\omega)$
are related with the generalized susceptibility of
the physical system $\alpha_x(\omega)$ by:
$$
\alpha_x(\omega)=\int_0^{\infty}\alpha_x(\tau)exp(i\omega t)d\tau
\equiv A^{-1}(\omega)
\eqno(3)
$$
with $\alpha_x(\tau)$ being the correspondent response function.
By using  the reciprocal Fourier transformation, from Eq. (3)
it is easy to show that $\alpha_x(\tau)$ satisfies the equation:
$$
A_t\{\alpha_x(\tau)\}=\delta(\tau)
\eqno(4)
$$
i.e., it describes the response of $x(t)$ at $\tau > 0$ induced by an 
impulsive perturbation at $\tau =0$.
\par
The mathematical formulation of the Onsager hypothesis,
i.e. the so called regression theorem  is that
the correlation function of fluctuations of the observable $x$, 
$C_{xx}(\tau)$, must satisfy at $\tau >0$  the same equation
as $\alpha_x(\tau)$ at $\tau >0$,
i.e. 
$$
A_t\{C_{xx}(\tau)\}=0 \ \ \ for \ \ \tau>0
\eqno(5)
$$
As a consequence of this theorem, 
the time dependence of both $\alpha_x(\tau)$ and $C_{xx}(\tau)$
is described by the same set of eigen-solutions
$exp(-i\omega_R t)$ obtained from the homogeneous equation:
$A_t\{exp(-i\omega_R t)\}=0$.
The resonant frequencies $\omega_R$ satisfy the dispersion
equation $A(\omega_R)=0$, i.e. in the complex frequency  plane
$\omega=\omega'+i\omega''$ 
they are  the poles of the susceptibility.
Due to the causality principle,
all poles of $\alpha_x(\omega)$ are placed in the lower half-plane [3], 
i.e. $\omega_R'' < 0$, what provides the relaxation process.

The response function formalism
allows us to describe macroscopic properties of the physical system
by considering merely the analytical properties of
$\alpha_x(\omega)$.
Let us consider the reasons leading to the violation of  QRT
by using the usual interrelation between the spectral density of the
fluctuating observable, $S_{xx}(\omega)$, and that of the 
force, $S_{ff}(\omega)$ [3],
$$
S_{xx}(\omega)=\alpha_x(\omega) \alpha_x(-\omega) S_{ff}(\omega)
\eqno(6)
$$
In the framework of linear response formalism, the 
correlation function of observable fluctuations  obeys the following
integro-differential equation: 
$$
A_t\{C_{xx}(\tau)\}={1\over 2\pi}\int_{-\infty}^{\infty}
\alpha_x(-\omega)S_{ff}(\omega)
exp(-i\omega \tau)d\omega
\eqno(7)$$
As follows from Eqs. (6) and (7), the fluctuations 
are governed by the properties of internal and external interactions.
The former is associated with the resonances of the susceptibility
$\alpha_x(\omega)$ and the latter with the frequency
dependence of $S_{ff}(\omega)$.
As follows from Eq. (7), the internal interactions can not be a source 
of violation of the regression hypothesis since,  
due to the causality principle,  $\alpha_x(\omega)$ has no poles in 
the upper half-plane, and, hence,
$\alpha_x(-\omega)$, has no poles in the lower half-plane.
Therefore, the right hand side of Eq. (7) is equal to zero for $\tau >0$
if  $S_{ff}(\omega)$ also has no poles in the lower half-plane.
Since the spectral density satisfies the condition
$S(-\omega')=S(\omega')$, its poles must be placed symmetrically
with respect to the real axis $\omega=\omega'$.
Therefore, the regression theorem can be fulfilled only in the case,
when $S_{ff}(\omega)$ is an analitycal function in the whole plane,
i.e. it has no poles anywhere.
\par
Usually, to determine the properties of $S_{ff}(\omega)$ under
thermal equilibrium one uses the QFDT which relates 
the spectral density $S_{xx}(\omega)$ with the imaginary part 
of the susceptibility, $Im\{ \alpha_x(\omega)\}$, by the simple relation:
$$
S_{xx}(\omega)=g(\omega)Im\{ \alpha_x(\omega)\}
\eqno(8)$$
where the relating factor 
$g(\omega)=\hbar coth({{\hbar\omega}/ {2kT}})$
is a universal function of frequency $\omega$ and temperature $T$ 
only.
Note, that the proof of the QFDT, given for the first time by
Callen and Welton [2] and in a somewhat other form in [3,4],
is strictly valid on the real axis, $\omega=\omega'$, only.
In applying the QFDT, the expression (8)
is usually extended to the whole frequency plane.
The imaginary part of $\alpha_x(\omega)$ has a natural
analytical continuation, defined as:
$Im\{ \alpha_x(\omega)\}=i[\alpha(-\omega)-\alpha(\omega)]/2$,
however the relating factor $g(\omega)$ is continued into the
complex plane as it is, without providing a sufficient physical justification.
By assuming that such a continuation is valid
and by comparing Eqs. (6) and (8), one obtains the main
relation which is used to determine the properties of  $S_{ff}(\omega)$
under thermal equilibrium: 
$$
S_{ff}(\omega)=g(\omega)Im\{A(\omega)\}
\eqno(9)
$$
where 
$Im\{A(\omega)\}=
Im\{\alpha_x(\omega)\}/[\alpha_x(\omega)\alpha_x(-\omega)]$
is the imaginary part of the operator $A_t\{\}$ written in the spectral
representation. 
This imaginary part is responsible for the relaxation processes
of fluctuations in the system under test 
(i.e. it determines the relaxation law)
and corresponds to the dynamical force of a friction written as 
$\tilde f_{\omega}=iIm\{A(\omega)\}x_{\omega}$.
As follows from Eq. (9), the analytical properties of 
$S_{ff}(\omega)$ are governed by two factors:
(i) the poles of  $g(\omega)$ coming from the QFDT, and 
(ii) the relaxation law of the system,
i.e. the frequency dependence of $Im\{A(\omega)\}$.
For the regression theorem to hold, it is sufficient
that the poles coming from $g(\omega)$ are compensated
by the zeros of  $Im\{A(\omega)\}$. 
For example, this compensation takes place
in the classical domain, where $\hbar \rightarrow 0$,
$g(\omega)\rightarrow {2kT/\omega }$, 
i.e. there is a pole at $\omega=0$.
To compensate this pole it is sufficient that 
$Im\{A(\omega)\}=\omega \zeta$,
where $\zeta$ is a constant independent of frequency.
This case corresponds to the usual viscous friction used
to describe the relaxation within the classical limit
when $S_{ff}(\omega)={2kT\zeta }$  
and the regression theorem is fulfilled.

In the quantum domain, where $\hbar\neq 0$,
but the relaxation is still described by the same classical law, when
$\hbar= 0$,
$S_{ff}(\omega)$ contains uncompensated
poles at the Matsubara frequencies
$\omega_m=i\Omega m$, where $\Omega=2\pi kT/\hbar$ and  
$m=\pm 1,\pm 2,...$
This is the source of  violation of the regression theorem, 
as demonstrated in [7] by calculating  the right hand side
of Eq. (7) for the damped oscillator.
However, in any case this result can not be considered
as a convincing proof of the QRT violation, since
a relaxation process within the quantum description does not necessarily
must  satisfy the same laws as within the classical description.
Thus, the validity of the QRT is directly related with the analytical
behavior of the relating factor, $g(\omega)$, in the QFDT.
Consideration of this problem is the task
of the following section.
\par
{\it The fluctuation-dissipation theorem.}
From the general principles of quantum statistics
the correlation function of fluctuations of an observable $x$ 
can be written in the operator representation as:
$$
 C_{xx}(\tau)={1\over 2}
Tr\{\hat \rho_S[\hat x(\tau)\hat x(0)+\hat x(0)\hat x(\tau)]\}
\eqno(10)
$$
while the corresponding function of the linear response 
is given by Kubo formula [3]:
$$
\alpha_x(\tau)={i\over \hbar}
\cases{
Tr\{\hat \rho_S[\hat x(\tau)\hat x(0)-\hat x(0)\hat x(\tau)]\}&,
  $\tau \ge 0$ \cr
0&,  $\tau < 0$ \cr }
\eqno(11)
$$
where $\hat \rho_S$ is the density operator which describes some
stationary state of the physical system under test characterized by
Hamiltonian $\hat H_S$ with a set of eigenstates $\psi_n$ 
corresponding to energy eigenvalues $E_n$.
The time dependence of both $C_{xx}(\tau)$ and  $\alpha_x(\tau)$
is determined by values of the operator $\hat x(t)$ taken
at two different time moments $t=0$ and $t=\tau$ only.
Therefore, in the matrix representation it is:
$x_{mn}(\tau)=x_{mn}(0)exp(-i\omega_{mn}\tau)$.
To obtain from Eqs. (10) and (11) the spectral dependences of
$S_{xx}(\omega)$ and $\alpha_x(\omega)$ applicable
in the whole complex plane, we make use of the 
standard  procedure of calculation of the Fourier transform in such a case: 
$$
\int_{-\infty}^{\infty} x_{mn}(\tau)exp(i\omega \tau) d\tau=
{1\over {i(\omega-\omega_{mn}-i\Delta')}}
-{1\over {i(\omega-\omega_{mn}+i\Delta'')}}
\equiv 
{2\Delta\over {(\omega-\omega_{mn})^2+\Delta^2}}
\eqno(12)
$$
Here the integration interval is subdivided into two parts:
$(-\infty,0)$ and $(0,\infty)$.
Inside the first and second part each of the Fourier integrals
converges separately to analytical functions of $\omega$ in
the lower ($\omega''<0$) and upper ($\omega''>0)$ half-planes,
respectively, and diverges in the opposite half-planes.
In the divergent regions the value of the integrals can be definded
only as analytical continuation of the expressions obtained in the
convergent regions.
The direction of the shift of the divergent poles is
determined by the sign of the $i\Delta$-terms in Eq. (12).
Under the usual assumption that $C_{xx}(\tau)=C_{xx}(-\tau)$,
such poles of $S_{xx}(\omega)$ are placed symmetrically with respect to
the real axis, i.e. $\Delta'=\Delta''=\Delta$.
When $\Delta \rightarrow 0$ the Fourier integral of Eq. (12)
tends to a $\delta$-function which we shall label as:
$2\pi \delta_{\Delta}(\omega-\omega_{mn})$,
where subindex $\Delta$ indicates that the analytical properties 
in the complex plane are determind in accordance with Eq. (12).
Since $\alpha_x(\tau)=0$ when $\tau < 0$,
the integral in Eq. (12) is calculated in the interval 
$(0,\infty)$ only, and hence, his value corresponds to the second term in 
Eq. (12).
\par
By going in Eqs. (10) and (11) to the matrix representation
and by applying the procedure of Fourier-integral calculations
given by Eq. (12) one obtains:
$$
\biggl [{ {S_{xx}(\omega)} \atop { \hbar Im\{\alpha_{x}(\omega)\} } }
\biggr ]
=
2\pi
\sum_{mn} |x_{mn}|^2
\biggl [ { {\rho_n + \rho_m} \atop {\rho_n - \rho_m } } \biggr ]
\delta_{\Delta}(\omega-\omega_{mn})
\eqno(13)
$$
where $\rho_n=<\psi_n|\hat \rho_S|\psi_n>$ is the probability
to find the system  in the eigenstate $\psi_n$.
As follows from Eq. (13), the analytical properties 
of both spectra $S_{xx}(\omega)$ and $\alpha_x(\omega)$
are determined by the same set  of symmetrical poles at
$\omega=\omega_{mn}\pm i\Delta$, each of which describes an allowed
transition between the eigenstates of the physical system.
The difference of each spectrum is that the weight represented by the terms 
($\rho_n \pm \rho_m$) with which a pole enters
into $S_{xx}(\omega)$ and $Im\{\alpha_x(\omega)\}$ is different,
i.e. both quantities differ in the value of the residue at the poles 
while the set of poles is the same.
Therefore, to interrelate  $S_{xx}(\omega)$ and $Im\{\alpha_x(\omega)\}$
(what is the content of QFDT) it is sufficient to formulate a procedure 
to recalculate their residues in each pole.
\par
Since at thermal equilibrium 
$\rho_m/\rho_n=exp(-\hbar\omega_{mn} /kT)$
is a universal function of the transition frequency $\omega_{mn}$ only,
the full set of all weight factors, 
which perform such a recalculation in the poles of Eq. (13),
can be described by a function of the current frequency given by:
$$
 g(\omega)=\hbar {1+p(\omega)\over 1-p(\omega)}
\eqno(14)$$
where it is  assumed that $p(\omega)|_{\omega=\omega_{mn}}=\rho_m/\rho_n$,
i.e. one obtains the expression of the relating factor in Eq. (8).
Such a recalculation of the residues,
is always present in more or less explicit form in any derivation of the QFDT.
Usually it is performed in the following way. 
The multipliers $1\pm p(\omega_{mn})$ are isolated from 
the double-sum sign in Eq. (13)
as functions of $\omega$, i.e.
$$
\biggl [{ {S_{xx}(\omega)} \atop { \hbar Im\{\alpha_{x}(\omega)\} } }
\biggr ]
=
2\pi
\biggl [ { {1 + p(\omega)} \atop {1 - p(\omega)} } \biggr ]
\sum_{mn} |x_{mn}|^2\rho_n
\delta_{\Delta}(\omega-\omega_{mn})
\eqno(15)
$$
Such a transformation is usually justified either 
by assuming a quasi-continuous spectrum of the physical system [2],
or by using the properties of the $\delta$-function 
inside the double-sum sign in Eq. (13) when $\Delta \rightarrow 0$ [3,4].
Then the double sum multiplier common to 
$S_{xx}(\omega)$ and $Im\{\alpha_x(\omega)\}$, 
which describes the resonances of the system under test,
is excluded from consideration.
As a result, one obtains QFDT given by Eq. (8)
where the spectrum of the resonances is not given in explicit form.
\par
At the level of Eqs. (10) to (15) there is no reasons
of QRT violation, since the properties of fluctuations and relaxation
are determined by the same set of resonances of the physical system.
The last step in the derivation of QFDT is the transition from Eq. (15) 
to Eq. (8).
Here a formal violation of the 
equivalence of such a transformation can take place when
$Im\{\alpha_x(\omega)\}$ has zeros at frequencies $\omega_m$,
where the multiplier $1-p(\omega_m)=0$.
Going from Eq. (15) to Eq. (8), the same multiplier $1-p(\omega)$
appears in the denominator of $g(\omega)$
so that  the r.h.s. of Eq. (8) will have
in hidden form uncertainties of ${0\over 0}$-type,
which can be a source of additional "properties" absent in Eq. (15).
Since in the original Eq. (15) $S_{xx}(\omega)$
is a regular function at frequencies $\omega=\omega_m$
it must remain a regular function in Eq. (8) too. 
This means that all poles of the relating factor $g(\omega)$ in QFDT,
and hence, the poles of $S_{ff}(\omega)$ (see Eq. (9)),
are points with removable divergence,
i.e.  the divergence at the poles must be compensated by
the zeros of $Im\{\alpha_x(\omega)\}$.
Thus, the correct solution of the uncertainty present 
in Eqs. (8) and (9)
means, in essence, the equivalence of QFDT and QRT 
from a physical point of view. 
\par
{\it Generalization of the fluctuation-dissipation theorem.}
Contrary to thermal equilibrium conditions,
in the general case $\rho_m/\rho_n$ is not a quantity
suitable to perform a recalculation of residues in the poles of 
$S_{xx}(\omega)$ and $Im\{\alpha_x(\omega)\}$.
Usually, physical systems are characterized 
by some set of transitions at the same frequency 
$\omega_{mn}=\omega_R$, as it takes place, for example,
in the case of an equidistant spectrum.
In this case $\rho_m/\rho_n$ loses its unique dependence on frequency and
can take different values for each separate transition.
Accordingly, in Eq. (13) we regroup  the double summation
over the states of the system in such a way to put
together all possible transitions with the same resonance
frequency $\omega_R$.
Then, we number the whole set of different resonant frequencies
in the whole frequency range $[-\infty ,\infty ]$, 
for example, 
by the subindex $R=0,\ \pm 1,\ \pm 2, ...$
so that $\omega_{R+1}>\omega_{R}$, where $\omega_R=\omega_{mn}$
is the resonant frequency of the system for all the transitions
which satisfy the condition $\omega_{mn}=\omega_{m'n'}$
at $n\neq n'$ and $m\neq m'$.
In so doing, it is asssumed that $\omega_R=\omega_{mn}$ and  
$\omega_{-R}=\omega_{nm}$ correspond to two different resonances
with frequencies related as $\omega_{-R}=-\omega_R$.
As a consequence, the double summation over the states in Eq. (13)
can be rewritten as:
$$
\sum_{m,n} |x_{mn}|^2\{\rho_n \pm \rho_m\}
\delta_{\Delta}(\omega_{mn}-\omega)=
$$
$$
\sum_R\delta_{\Delta}(\omega_R-\omega)
\left[
\sum_{E_m=En+\hbar\omega_R}'\rho_n|x_{mn}|^2
\pm \sum_{E_n=Em-\hbar\omega_R}'\rho_m|x_{nm}|^2
\right]
\eqno(16)$$
Here the prime sign over the sum symbol inside the square brackets
means that, when summing over the states
the only contribution comes from  $n$ and $m$ states
satisfying the conditions reported at the bottom of the sum symbol.
The first and second prime-sum in Eq. (16) at $\omega_R>0$
collect all the transitions accompanied, respectively, by an increase
and a decrease of the system energy by $\hbar\omega_R$.
Accordingly, when $\omega_R<0$ they  interchange their places.
It is easy to see that by defining the function $p(\omega)$ as:
$$
p(\omega)\big|_{\omega=\omega_R}=
={{\sum_{E_n=Em-\hbar\omega_R}'\rho_m|x_{nm}|^2}
/ {\sum_{E_m=En+\hbar\omega_R}'\rho_n|x_{mn}|^2}}
 \eqno(17)$$
Eq. (16) can be written in a form analogous to that of Eq. (15),
where $p(\omega)$ satisfies the condition:
$p(-\omega)=p^{-1}(\omega)$ 
and is uniquely defined in the whole set of resonant frequencies,
i.e. at $\omega=\omega_R$.
Thus, it is always possible to write 
a generalized QFDT in terms of Eq. (8) 
where the relating factor can be represented in the sufficiently
general form given by Eq. (14)
and $p(\omega)$ is given by Eq. (17).
\par
{\it In conclusion,}  as a consequence of the nonequivalent
transformation used to derive the QFDT [2-4],
it is shown that the QFDT given by Eq. (8) contains an uncertainty
of ${0\over 0}$-type originated by 
the poles of the relating factor $g(\omega)$
and the zeros of the imaginary part of the susceptibility 
at the so called Matsubara frequencies.
A correct resolution  of such an uncertainty 
means, from the physical point of view, 
that the spectral properties of both
fluctuations and relaxation are driven by the same set
of resonances of the physical system under test only.
In the framework of such a formulation,
the QFDT is equivalent to QRT
and the former can be considered as a mathematical representation
of the Onsager hypothesis in the frequency domain.
Moreover, a relation analogous to the QFDT
can be written for an arbitrary stationary state of the physical system,
by supposing that its linear response satisfies the causality principle.
\par
In the framework of the Langevin approach, the QRT is fulfilled
under the requirement that
the spectral density of the force $S_{ff}(\omega)$,
which drives the fluctuations of observables, is an analytic function 
in the whole complex plane.
Since the causality principle simply claims that
there must exist some relaxation process in the physical system,
but it does not detail the relaxation law,
this requirement can be considered as a kind of selection rule
for the choice of the relaxation law of fluctuations
in order to be in agreement with the QRT.
It is evident that such a requirement  will call for 
some revision of the Langevin approach in the quantum domain.
\par
{\it Acknoledgments.} This work has been supported by:
{\it Cooperation franco lituanienne Projet 5380} of French CNRS,
the high-level grant DRB4/MDL/no 99-30 of the french 
{\it Ministere de l'Education
nationale, de la recherche et de la technologie}, 
NATO collaborative-linkage grant PST.CLG.976340 
and the Galileo project n. 99055.
\vskip 1.2truecm\noindent
\centerline{\bf References}
\item {1.}
L. Onsager, {\it Phys. Rev.} {\bf 38}, 2265 (1931).
\item {2.}
H.B. Callen and T.A. Welton, {\it Phys. Rev.} {\bf 83}, 34 (1951).
\item {3.}
L. Landau and E. Lifshitz, {\it Statistical Physics}
(Addison Wesley Reading, Mass., 1974)
\item {4.}
Sh. Kogan, {\it Electronic noise and fluctuations in solids}, Cambridge
University Press (1996).
\item {5.}
H. Grobert, {\it Z. Phys.} {\bf B 49}, 161 (1982).
\item {6.}
P. Tolkner,  {\it Ann. Phys. (N.Y.)} {\bf 167}, 380 (1986).
\item {7.}
G.W. Ford and R.F. O'Connell, {\it Phys. Rev. Lett.} {\bf 77}, 
798 (1996).
\item {8.}
M. Lax,  {\it Phys. Rev.} {\bf 129}, 2342 (1963).
\item {9.}
V.H. Louisell, {\it Quantum Statistical Properties of Radiation}
(Wiley, New York, 1973), Sec. 6.6.
\item {10.}
C. Cohen-Tannoudji, J. Dupont-Roc and G. Grinberg,
{\it Atom-Photon Interactions} (Wiley, New York, 1992), 
Chap. IV, p. 350.
\item {11.}
A. Mandel and E. Wolf, 
{\it Optical Coherence and Quantum Optics}
(Cambridge University Press, Cambridge, 1995), Sec. 17.1.

\bye